# Learning Design Preferences
# through Design Feature Extraction and Weighted Ensemble

Dongju Shin[1,2], Sunghee Lee[2], and Namwoo Kang[1,2,*]

[1] Cho Chun Shik Graduate School of Mobility, KAIST

[2] Narnia Labs

*Corresponding author: nwkang@kaist.ac.kr

**Abstract**

Design is a factor that plays an important role in consumer purchase decisions. As the need for understanding and predicting various preferences for each customer increases along with the importance of mass customization, predicting individual design preferences has become a critical factor in product development. However, current methods for predicting design preferences have some limitations. Product design involves a vast amount of high-dimensional information, and personal design preference is a complex and heterogeneous area of emotion unique to each individual. To address these challenges, we propose an approach that utilizes dimensionality reduction model to transform design samples into low-dimensional feature vectors, enabling us to extract the key representational features of each design. For preference prediction models using feature vectors, by referring to the design preference tendencies of others, we can predict the individual-level design preferences more accurately. Our proposed framework overcomes the limitations of traditional methods to determine design preferences, allowing us to accurately identify design features and predict individual preferences for specific products. Through this framework, we can improve the effectiveness of product development and create personalized product recommendations that cater to the unique needs of each consumer.

Keywords: Design Preference Learning, Heterogeneous Preference, Design Recommendation, Deep Learning, Feature Extraction, Wheel Design

# 1. Introduction

The appearance of a product is significant, creating a first impression on the customer's mind, it has been widely recognized as a crucial factor in consumer decision-making. Over the years, several studies have established that a product's design is one of the key factors that influence customer decision-making (Kotler and Rath, 1984; Bloch, 1995; Veryzer and Hutchinson, 1998; Bloch et al., 2003; Pan et al., 2017). According to Bloch (1995), the design shape of a product plays a crucial role in capturing the customer's attention, conveying product information, and providing a visually satisfying experience, which in turn creates a long-lasting perception of the product. In practice, sales predictions can be significantly improved by adjusting designs (Landwehr et al., 2011).

In particular, with the advent of the 4th industrial revolution, mass customization became in the spotlight, and the design of products became more important. The transition from mass production through previous automated production processes to personalized production manufacturing based on artificial intelligence (AI) is taking place, and companies should be able to identify various needs of consumers and provide products that reflect individual needs (Liu et al., 2021). According to a study by Deloitte (Fenech and Perkins, 2019), in some categories, more than 50% of consumers expressed interest in purchasing customized products or services, with the majority of them willing to pay more for a customized product or service. Furthermore, Liu et al. (2017) analyzed 202 car models sold in the United States from 2003 to 2010, and proved that aesthetic design has a significant impact on consumer preferences. Therefore, to ensure that a product succeeds in the market, it is essential to understand the customer's needs and preferences and designing a product that reflects what customers prefer. It shows us that when designing a product, it's important to anticipate customer preferences for design.

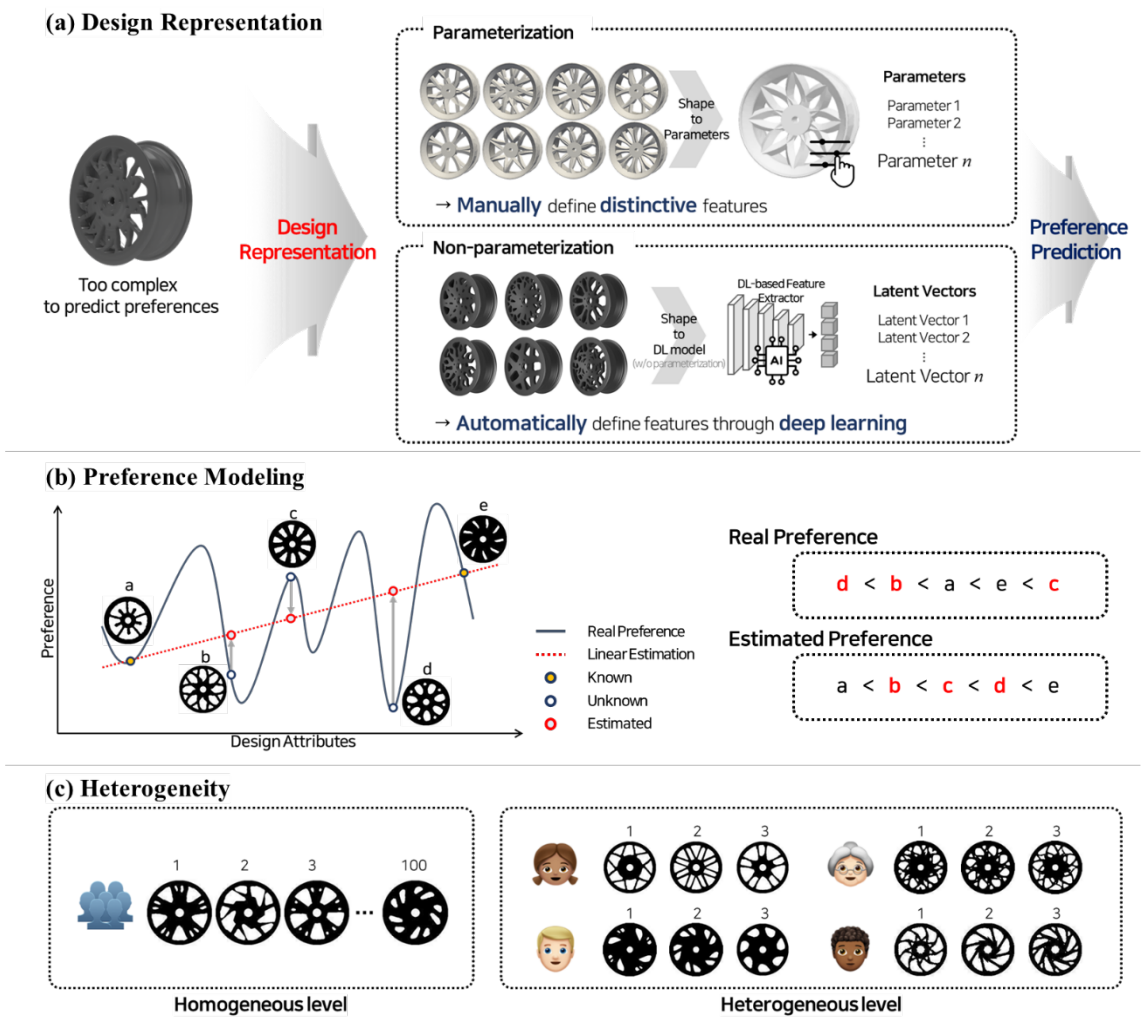

**Figure 1 Three main considerations for preference prediction**

There are three important factors to focus on when predicting customer preferences: Design Representation, Preference Function, and Heterogeneity (see **Figure 1**).

**Design Representation**: To create a design for evaluation or to evaluate a design made, the design must be expressed in several parameters. Design parameters include size, weight, and dimensions of a particular part, etc., but only a few parameters are not enough to capture all the complex and aesthetic designs. Recent advances in Deep-Learning (DL) have made it possible to extract features of high-dimensional image data nonlinearly without parameterization and express them in low-dimension (e.g., Auto Encoder(AE), Variational Auto Encoder (VAE), Generative Adversarial Networks (GAN)). However, these methodologies only find features from the perspective of computer matrix and cannot capture the essential characteristics of design that designers focus on (**Figure 1a**).

**Preference Modeling**: Preferences are approximated in most cases based on survey data. existing methodologies mainly utilize simple Machine-Learning (ML) methods such as linear, quadratic, and Hierarchical Bayesian (HB), etc., and some of them also utilize DL such as neural network to fit preference prediction models. ML methods have limitations for the highly non-linear preferences. Preference is an unquantifiable area, it can be approximated numerically as a function that is very nonlinear for the software to recognize. Using DL can partially solve the non-linearity problem, but it requires a large amount of survey data (**Figure 1b**).

**Heterogeneity**: People's tastes in design are highly individualized and differ greatly from one another. In order for a product to succeed in marketing, it is necessary to select a target group or customer well and then create a product that the target wants. Given the subjective nature of preferences, it can be challenging to design products that appeal to everyone. It is therefore crucial to comprehend these individual differences to design products that meet the expectations of consumers. For this reason, scholars construct preference models of individual-level rather than population-level (Orsborn et al., 2009; Pan et al., 2017; Dotson et al., 2019; Kang et al., 2019; Burnap et al., 2023), considering different tastes of consumers. For individual-level preference modeling, a lot of preference data from a single individual is required, but it is difficult to collect enough data from a survey to train a DL model on a single person. As a result, training an individual model with a small amount of data can result in a highly biased model (**Figure 1c**).

In order to predict preferences, reasonable method is needed from all three perspectives. The objective of this research is to build a heterogeneous design preference prediction model through design features that reflect designers' domain knowledge. For this purpose, this research concentrates on defining design features that well capture the design itself and the design parameters that human defined, and training a preference prediction model that well understands human preference trends from the restricted and sparse individual datasets. Finally, this research develops a design recommendation system that can recommend reasonable designs for target customers or groups by predicting preferences for new designs.

These objectives lead to following research questions:

**Research Question**: How can we better learn individual preferences for complex designs that are difficult for humans to parameterize?

**Problem 1**: DL can be used to extract features of complex designs without parameterization, but it cannot reflect the designer's domain knowledge.

**Problem 2**: Learning individual preferences for complex designs requires a lot of data.

To address those two problems, we conduct following tasks: For Problem 1, we propose a method to obtain features that simultaneously reflect the designs and the designer's domain knowledge. This is a non-parameterization data representation method that does not involve manual parameterization, but still takes into account human-defined parameters. For Problem 2, we propose an ensemble based on individual preference similarity that can improve predictive power with less data. DL based utility model is defined using Siamese network and weighted ensemble technique to predicts heterogeneous preferences while considering preference tendencies of others. Finally, a design recommendation system is suggested, with a focus on 2D wheel design as an example.

Through proposed framework, we can predict individual design preferences and design preferences for target market customers. it can assist in the analysis of the target market and aid in product design decisions. This framework has the potential to expedite mass customization through the application of AI, providing companies with valuable insights into customer preferences.

The paper is structured as follows: Section 2 reviews the related works of ML-based preference modeling method. In Section 3, we present our proposed framework with three stages: design representation, heterogeneous preference modeling, and design recommendation system. Section 4 describes results of the experiments on the model performances and discuss about the result of preference prediction. Finally, Section 5 concludes our research with the contributions and limitations.

## 2. Literature Review

In this section, we introduce related works to understand this study, mainly, ML/DL-based design preference models in relation to design preferences.

### 2.1 ML and DL Based Preference Modeling

**Table 1 ML and DL based preference modeling studies**

| Research | Design Represen-tation | Design Representation method | Preference Function | Hetero-geneity | Application | Survey type |
|---|---|---|---|---|---|---|
| Orsborn et al., 2009 | Parameterization | Vectors, Scalars, Angles of Curves | Quadratic | O | 2D SUV Sillhouette | Choice |
| Tseng et al., 2012 | Parameterization | eight cubic Bezier curves | Neural network | X | 2D Vehicle sillhouette | Rating |
| Reid et al., 2013 | Parameterization | Scalar of Height, Width, Curvature, etc. | Linear | X | Vehicle (Sillhouette, Rendering) Carafe (Sketch, Rendering) | Choice |
| Dotson et al., 2019 | Parameterization | A vector of 343 key points in 2-D space | Linear | O | 2D Vehicle Sillhouette | Choice + Rating |
| Kang et al., 2019 | Parameterization | 26 control points of Bezier curve | Rank SVM mix and HB | O | 3D Vehicle Rendering | choice |
| Pan et al., 2017 | Non Parameterization | cGAN (1024) | Siamese network | O | 2D SUV Image | Ranking |
| Sisodia et al., 2022 | Parameterization + Non Parameterization | VAE | - | - | 2D Watch Image | - |
| Burnap et al., 2023 | Non Parameterization | VAE-GAN (512) | Neural network | X | 2D SUV/CUV Image | Rating |
| This work | Parameterization + Non Parameterization | Multi-modal VAE (10 dim) | Siamese network | O | 2D Wheel Image | Ranking |

AI has become as a state-of-art approach and gained popularity for its ability to effectively process and analyze extensive datasets. As it turns out that customer preference is an important factor in products but hard to quantify, there are attempts to approximate design preferences by applying ML/DL (see **Table 1**).

**2.1.1 Design Representation by Parameterization**

Reid et al. (2013) designed a vehicle shape, Lugo et al. (2012) a wheel rim, both used similar techniques, employing a linear preference function determined through regression analysis. But they defined product attributes manually and considered homogeneous level preference. Sylcott et al. (2013) incorporated interaction terms with linear preference function, but also using aggregated data. On the other hand, Orsborn et al. (2009) used quadratic function to approximate individual-level choice-based utility and Kelly et al. (2011) enabled both quadratic preferences and potential interactions for homogeneous inferences.

A few studies combined design and performance (Dotson et al., 2019; Kang et al., 2019). Tseng et al. (2012) have developed a method for integrating the interconnection of design and functional constraints in computational design, despite their apparently dissimilar. An artificial neural network (ANN) model was developed to quantify surveyed consumer judgments of stylistic form. Combining this ANN model of design with a genetic algorithm (GA) enables concurrently accounting for multiple objectives in the domains of design and more functional performance evaluation. ANN approximates respondents' feelings towards vehicle silhouette designs generated through parametric design. However, there are still limitations in that generative design is carried out with a few simple parameters, only inference of already defined design attributes is possible, and it is a model at the homogeneous level.

Dotson et al. (2019) distinguished between a respondent's evaluation of a design and their overall evaluation of a product and its other features by using vehicle silhouette images. They enhanced the standard CBC task by adding direct ratings of image appeal to account for the utility correlation between images that look similar. They also modeled heterogeneity in image appeal using a multinomial probit approach, where the utility error covariance depended on the correlations in consumer image ratings.

Similar to Dotson et al. (2019), Kang et al. (2019) considered both design and function. They optimize the aesthetic design of vehicles to suit both form and function. When conducting a survey consisting of interactive bi-level questions, only form is provided for odd questions and form and function information are provided for even questions. The 19 design parameters used in 3D rendering of a vehicle are converted into 26 control points based on Bezier curve, allowing the car to change shape in real time. Based on the respondents' answers, they generate questions in real time using the rank SVM mix (Evgeniou et al., 2005), and after the survey, they use HB to create a heterogeneous overall preference model to check the hit rate for the verification questions. Through this, it is possible to learn the design and attributes preferred by consumers together, and to compare them in consideration of the balance of all factors. It provides an optimized form for consumer groups and analyzes sensitivity to functional elements. However, 19 design parameters do not contain enough aesthetic elements to express 3d features, and because they generate and train the choice of respondents, optimizing in the wrong direction can lead to undesirable designs. All these methods mentioned above were also constrained to manual design parameterization. It is useful for users who need to focus on some parameters that are meaningful, but the parameters are defined manually and a limit to scalability for other features to be considered exists.

**2.1.2 Design Representation by Non-Parameterization**
On the other hand, Pan et al. (2017) and Burnap et al. (2023) used data-driven design feature extraction, without parameterization. Pan et al. (2017) proposed Deep design, which predicts established the aesthetic attributes for vehicles. They trained cGAN with the design labels (release year, brand, model name, vehicle type, view point, color) as conditions for the vehicle, and formulated a discriminator that captures the link between the design images and the design labels. After that, a survey is conducted with an SUV image to have the respondents rank four vehicle images of the same view point according to the given attributes (e.g. sporty vs. conservative). A siamese network with two subnetworks is constructed using part of the discriminator of the trained cGAN, where the design labels and customer labels are put as input to predict the customer's aesthetic perception for each attribute. After that, the saliency map is extracted through guided backpropagation to visualize the important region of image for model prediction. This study can capture the heterogeneity of the market to see how design is perceived in the target market, but similar to the manual parametrization methods, it used pre-defined design labels to search features. It is difficult to expand on new designs or markets with undefined labels because it is trained with limited and pre-defined attributes, design labels, and customer labels.

Sisodia et al. (2022) proposed a non-parametrization method with design attributes as parameters to define product attributes that influence human preferences for unknown products. They combined non-paramertazation and paramerterization to learn a disentangled latent space so that products can be sorted for user-desired attributes. However, this work is limited by the fact that the parameter used as a label is a single product attribute, and they did not verify that the defined latent space really affects preference prediction. In contrast, Gabel and Timoshenko (2022) proposed a deep learning-based product choice model that considers consumers' purchase patterns, but did not include consideration of product attributes.

Burnap et al. (2023) addressed the delicate integration between machine intelligence and existing human workflows. Through the VAE-GAN model, their framework is to predict the aesthetic rating of vehicle design and generate new designs that fit the desired concept. Data are collected by conducting a rating survey on SUV/CUV vehicle images. The VAE encoder is used for transfer learning (TL), and the embedded latent vector is used to predict the rating score through the predictive model. The generative model is trained by utilizing the distribution of VAE that has trained vehicle image data. Design attributes are input into the generative model to generate a new image of the corresponding attributes. It is meaningful in that proposed VAE-GAN can predict consumer aesthetic judgments and create novel designs at the same time. Still, there are some limitations that it was a homogeneous level model and new attributes cannot be considered because limited design attributes have been defined. These data-driven methods are useful for design feature extraction without actual design parameters, but can't consider the parameters that are meaningful.

Lee and Kang (2020), a preceding research of ours, predicted individual preference with Siamese network. It is meaningful for introduce method for individual preference learning. But they used data-driven method for defining design features, not considering designers' domain knowledge, and also used ensemble with manual searched weights only considering individuals and other people, not considering preference similarity of people or weights training. We propose a framework to solve these problems. We solve the high-dimensional nature of the data by embedding a with a well-understood dimensionality reduction model. This complement the limitations of both parameterization and non-parameterization, by training models with the parameter that reflect designers' domain knowledge, but allowing designs to be extracted without manual intervention. To approximate individual heterogeneous preferences, heterogeneous preference models were constructed. It is possible to analyze human preferences and differences in preference designs by taste without defining categories for styles.

## 3. Methodology

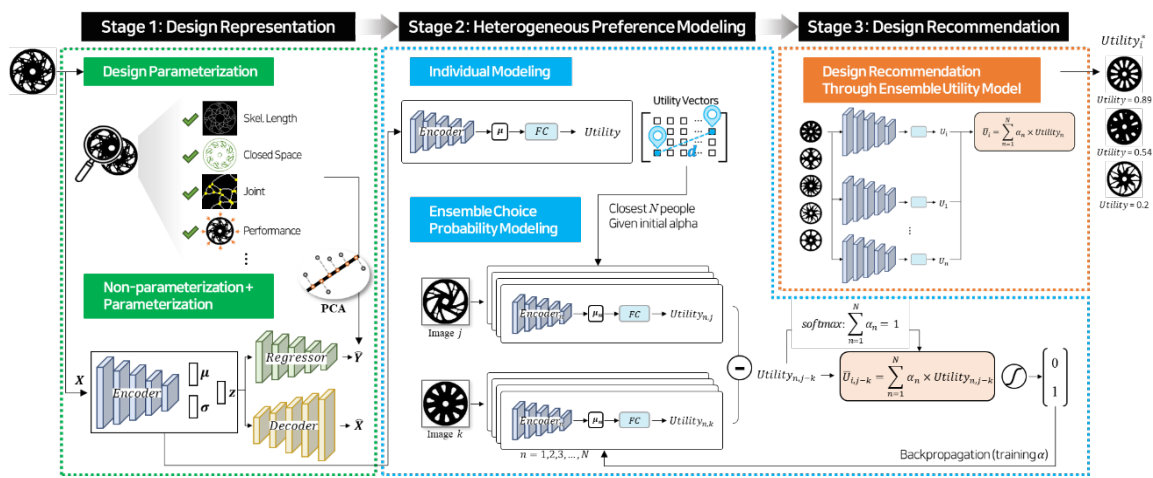

**Figure 2 Overall research framework**

The wheel design preference prediction and recommendation system proposed in this study is divided into three stages and follows the framework as shown in the **Figure 2**: Design Representation, Heterogeneous Preference Modeling, and Design Recommendation.

In stage 1, we establish DL-based dimensionality reduction model, called multi-modal VAE. In order to encode the high-dimensional wheel shape into a low-dimensional latent space, the DL model should understand the shape of the wheel and also the characteristic impression that human cognize, or designers have defined together. Multi-modal VAE can encode design shape into representational features while understanding both the shape and the characteristic features.

In stage 2, we train a DL model that predicts a user's choice probability by using two wheel images as inputs. The training data was obtained through an online survey of customers. For training, we use a Siamese network that shares model weights for multiple inputs to predict a single value of which the user will choose. we proceed TL with the encoder of multi-modal VAE. Ensemble technique is used to reflect preference tendency of similar people to the model.

In stage 3, recommendation system can be built by the choice probability prediction model in stage 2. Since the choice probability prediction model has binary label data, by removing the sigmoid layer at the end of the model, we can get user's utility of the design. Wheel designs with high utility can be recommended individually.

### 3.1 Design Representation

Designers have traditionally employed design parameters to represent the product shape, but have struggled to incorporate complicated shapes into overall a few parametric values. Even if latent vector that reconstruct a shape well is trained, it cannot be guaranteed that the latent vector is a feature that captures a characteristic impression that humans feel. To overcome this challenges, Multi-modal VAEs aims to create a latent space that better approximates the characteristic features of the design. In this context, the role of encoder is significant. Its primary function is to create latent vectors that represent the designs. To ensure that the encoder generates high-quality latent vectors that can capture the shape and the characteristic impression of the design, we define parameters that can contain the characteristics of each design. Subsequently, the decoder and regressor are connected to the encoder in parallel, allowing the latent vector to reconstruct the design while approximating the defined parameters.

#### 3.1.1 Data Generation

For the training data, 2D wheel image is generated by Deep generative design proposed by Oh et al. (2019). Label data is gained by feature engineering and transformed into low-dimensional space by PCA.

*a) Topology Optimization*

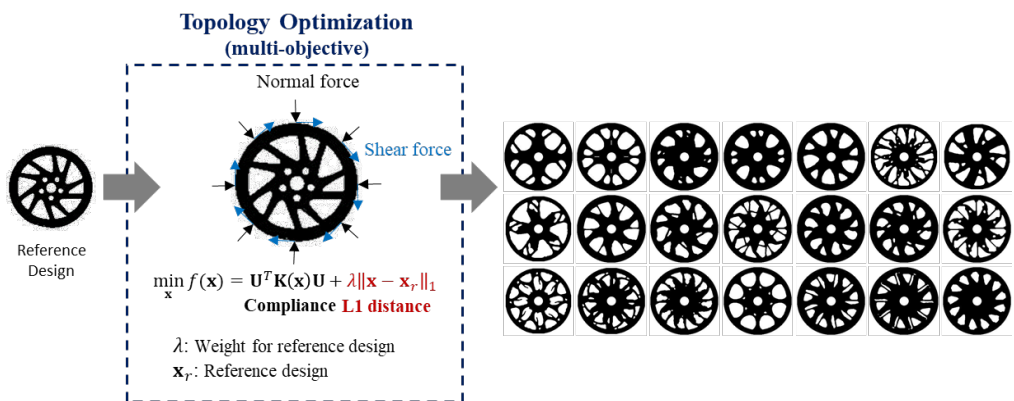

**Figure 3 Generative design by topology optimization (Oh et al. 2019)**

We generated wheels by generative design using topology optimization (Oh et al., 2019). The wheel generation process is shown in the **Figure 3**. Wheel designs collected by web crawling are preprocessed with binary images, which are referred to as reference images. By conducting TO, 2d design domain is optimized regarding the reference image and design parameters inputted by user. Multi-objective TO is conducted, minimizing compliance is the first objective function, and converging the L1 norm of the optimizing shape and reference image to the target similarity is the second objective function. To diversify the wheel designs, user can verify design parameters (reference image, target similarity, load ratio, volume ratio, number of symmetries). The target similarity determines how much the optimized shape will be similar to the reference shape. As the target similarity increases, the optimized image resembles the reference image, and as it approaches 0, the reference image is ignored. The load ratio refers to the ratio of the size of the uniform normal force to the uniform shear force over the rim section of the wheel. As the ratio of the normal force increases, a straight spoke shape is generated, and as the ratio of the shear force increases, a rotating spoke shape, like pinwheel, is generated. The volume ratio refers to the threshold of volume constraint. The number of symmetries determines how many symmetries the optimized wheel should be made (e.g. if the number of symmetries is 5, the design domain to optimize is a sector with a central angle of 72 degrees. After TO, the optimized sector is rotated 5 times to compose a circle). By combining various reference images and design parameters, millions of wheels can be generated that is meaningful from an engineering perspective. Multi-modal VAE and the preference prediction model were trained with 1,156 generated wheel data, augmenting them 10 times by rotating.

*b) Design Parameterization*

The design may be defined by various parameters. We classified wheel characteristics into morphology perspective, performance perspective, and others. A total of 11 parameters were acquired, and the parameters from each perspective are shown in **Table 2**.

**Table 2 Parameter definition for feature engineering**

| Perspective | Parameters |
| --- | --- |
| Morphology | Number of wheel symmetry, Length of a skeleton, Number of closed spaces, Number of joints, Sum of joint branches |
| Performance | Compliance for normal force, Compliance for shear force, Volume fractions of 2D wheels, Rim stiffness, Disk stiffness, Weight |

*b1) Morphology Perspective*

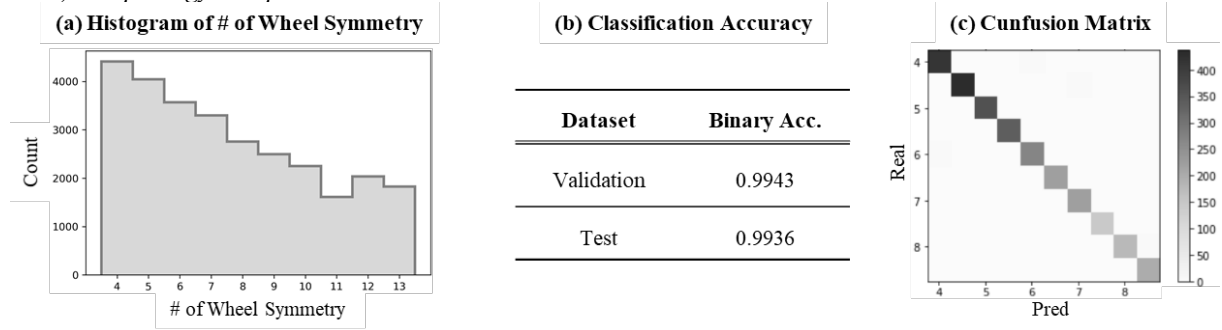

**Figure 4 Classification model for wheel symmetry**

To obtain number of wheel symmetry, we trained 2d cnn model. We used 28,293 generated TO data as input, and the label was a symmetry condition given when TO conducted (between 4 - 13). **Figure 4a** shows the number of data for each label. This 2d CNN model solves a classification problem with 10 classes, using softmax. **Figure 4b** and **c** is a training result. The accuracy for the validation, and test datasets was 0.9943 and 0.9936, respectively, showing very high accuracy, and we can see that the model has been trained to match labels for all classes. By confusion matrix. We utilize this model to get the number of wheel symmetry.

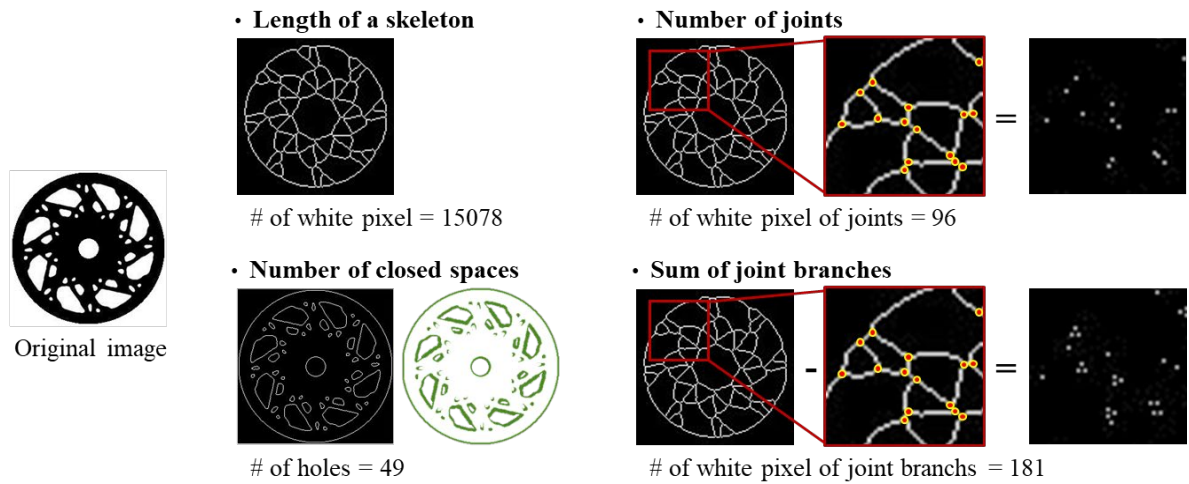

**Figure 5 Morphology parameters**

To obtain other morphology parameters (the length of a skeleton, number of closed spaces, number of joints, sum of joint branches), we refered to Ryu et al. (2021), using skeletonized parameters (see **Figure 5**). We tried to capture the complexity of the design, or ornateness, by extracting characteristics of the skeleton that we could get from the design. MATLAB image processing toolbox is used for skeletonization and extracting characteristics of the skeleton.

- Length of a skeleton: Sum of total pixels in the skeleton image
- Number of closed spaces: Number of closed holes of edge image
- Number of joints: Skeleton Number of points where branches of the Skeleton image intersect
- Sum of joint branches: Sum of the branches intersecting at joint points

*b2) Performance Perspective*

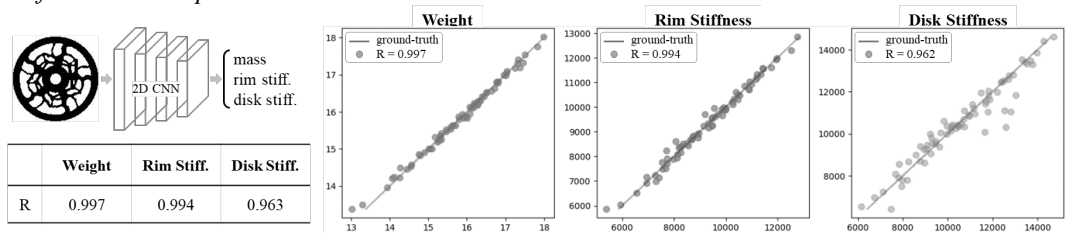

**Figure 6 Prediction model for engineerng performance**

We utilized global performance values as parameters that can affect geometry changes. The calculation of the compliance for each normal force and shear force was referred to Oh et al. (2019), calculated on 2d wheel images. Volume fractions of 2D wheels was calculated when the full circle inside the border of rim was set at 1.

For the rim stiffness, disk stiffness, and weight, 2D CNN model was trained to approximate the performance parameters. we utilized Yoo et al. (2021), generate 3d wheels, combined 2d wheel images with a single rim cross-sectional image. With the generated 3D wheel, modal analysis was carried out and gained label data for 2d CNN model. For training, 2d wheel image was used as an input and rim stiffness, disk stiffness, weight values were used as labels for regression problem. Although, because of the limitation for the automation of modal analysis, 722 data were gained and used to train the model. The 2d CNN regression model to predict performance had a good performance (see **Figure 6**).

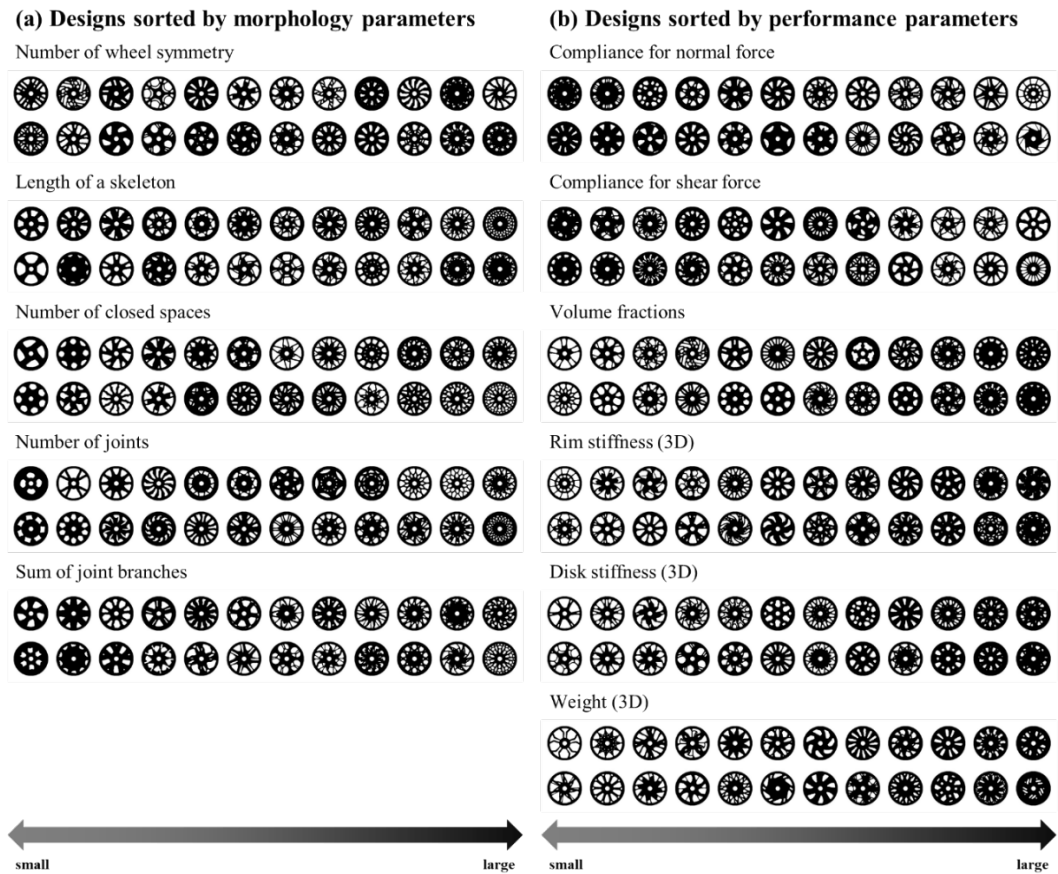

**Figure 7** Wheel designs sorted by parameters

To verify that the extracted parameters do indeed affect the design changes, we listed 24 images selected at regular intervals for each parameter (see **Figure 7**). First, the number of wheel symmetry aligns wheels with more spokes as the value increases. The other morphology parameters followed a similar trend, with more complex shapes being aligned as the value increased. Since morphology parameters were extracted based on skeleton, the weight of the wheel was not considered, and designs were sorted based on the complexity of skeletons. Performance parameters could address this problem. For the compliance for normal force, the more spoked, straight, and thick the designs were sorted for the smaller the compliance, while the more pinwheel-like rotating and spoke-thin designs were sorted for the larger the compliance. Conversely, when sorting by the compliance for shear force, smaller compliance tends to show more rotating spokes, while larger compliance tends to show straighter and thinner spokes. This is consistent with the physical concept that straight spokes are stronger for normal forces and rotating spokes are stronger for shear forces. For the other performance parameters, the volume tended to increase with increasing parameter values, as expected.

*b3) PCA*

After collecting all 11 parameters, it should be preprocessed to be a label data. Among the 11 parameters, there may be parameters that greatly affects the shape and parameters that does not. In addition, there may be parameters that needs to be deleted due to a strong correlation with other parameters. However, we found it difficult to quantitatively identify which parameters is of great importance to represent high-dimensional designs. Therefore, we utilized Principal Component Analysis (PCA), which can reduce the dimension of the data while preserving the distribution of the original data as much as possible (Dunteman, 1989). PCA combines existing features to create new features, that is, Principal Components (PCs). (e.g. The first main component PC1 preserves the distribution of the original data the most, and the second main component PC2 preserves the distribution of the original data the most.)

In this study, 11 parameters were transformed to 4 PCs, and explained variation ratio was 0.5234, 0.2650, 0.1823, and 0.0116, respectively. This means that 52.34% of the original dataset variance lies on the PC1 axis, and 26.50%, 18.23% lies on the PC2 and PC3 axis, respectively. PC4 axis contains a very small amount of information, about 1.16%. With four PCs, the sum of the variances is 0.9823, which results in about 2% loss of information in the variance of the original dataset, but it can be seen that PCA encoded 11 parameters well enough. We proceed with min-max scaling twice. First, in the transformation to PCA, the scales of each of the 11 parameters are very diverse, so each is scaled to a value between 0-1 through minmax scaling. minmax scaling is performed even after converting to PCs, so that it is used as a label data. This is to adjust the label value to 0-1 again so that the regressor can be trained well.

### 3.1.2 Multi-modal VAE

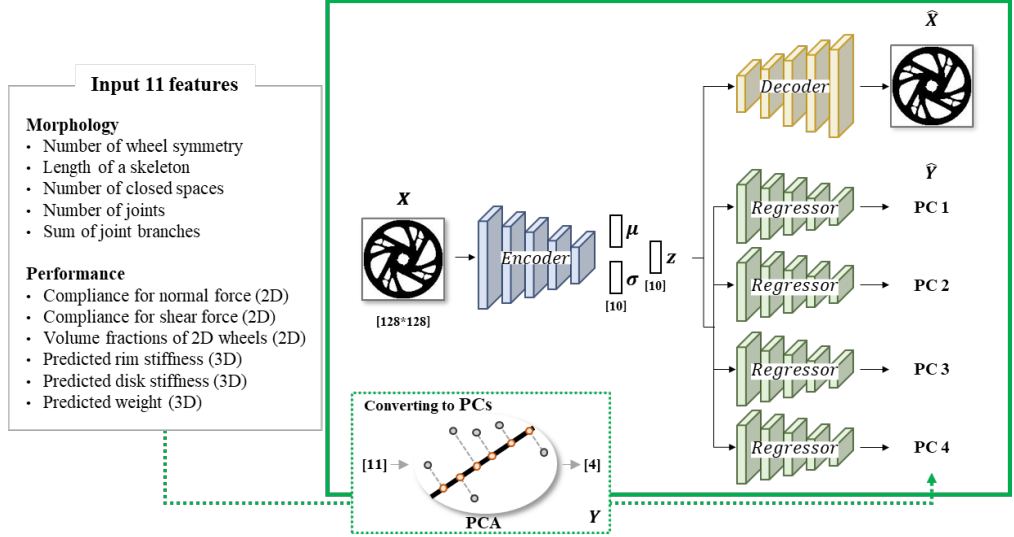

**Figure 8 Multi-modal VAE architecture**

In this study, model was trained to improve the quality of latent space, which enables to replace the design attributes with the latent vectors. In the latent space, similar designs may be located at close distances, and that designs that are different from each other are located at far distances. The parameters extracted through feature engineering are used to do so, and the architecture is shown in the **Figure 8**. In the same way as the convolutional VAE, the image is encoded through encoder and reconstructed through decoder. The regressor was added in parallel with the decoder so that the latent vector pass through the regressor and approximates the 4-dimensional PCs.

We used combined loss function for two networks; VAE loss, Regressor loss:

$$\alpha_1 (\sum_{j=1}^{D} x_{i,j} + log p_{i,j}(1-x_i) log(1-P_{i,j}) - \frac{1}{2}\sum_{i=1}^{J}(\mu_{i,j}{}^2 + \sigma_{i,j}{}^2 - \ln(\sigma_{i,j}{}^2) - 1)^2) + \alpha_2 \frac{1}{n}\sum_{i=1}^{n}(C_i - \widetilde{C_i})^2. \quad (1)$$

VAE loss uses BCE to reduce the difference between the reconstructed image and the input image, and KL divergence to ensure that the distribution of the sampled latent vector follows the distribution of the existing latent space. Regressor loss is defined as MSE loss between the 4 true and predicted values. We weight sum those two loss functions ($\alpha_1$ for VAE loss, $\alpha_2$ for regressor loss in Eq 1. $\alpha_1, \alpha_2$ were set to 1, 10, respectively.).

## 3.2 Heterogeneous Preference Modeling

The preference prediction model was trained heterogeneously depending on the person. This borrowed the concept of the utility model. Creating a single utility model was not appropriate because everyone has individual, non-linear preferences.

### 3.2.1 Data Generation

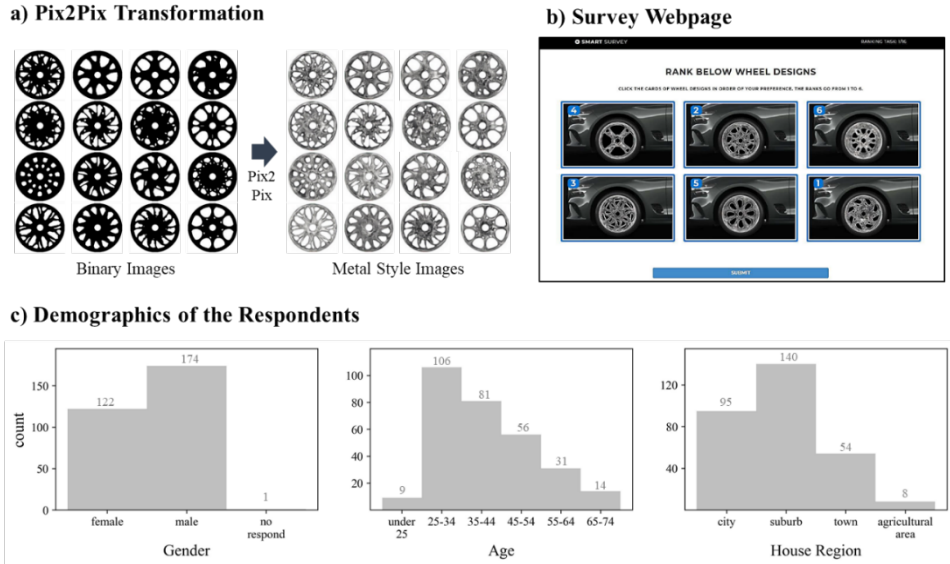

**Figure 9 Customer survey (Lee and Kang, 2020)**

We used customers' survey data from Lee and Kang (2020). They conducted a survey by synthesizing metal-style wheel images on top of a vehicle image to select wheels that fit the vehicle. Pix2pix is used to convert binary wheel images to metal-style wheel images reflecting the material of the actual wheel (**Figure 9a**). The survey was conducted by ranking six images (**Figure 9b**), and each respondent will solve a total of 16 questions, which means that 96 wheels will be evaluated by a respondent. They used Latin Hypercube Sampling (LHS) to allow various wheel designs to be composed of one problem, at which time the 7th and 8th problems were equally given to all respondents so that test dataset can be unified. The survey was conducted among people living in the U.S in Mturk (Mturk, 2020), a total of 300 people participated in the survey, and 297 survey data that removed outliers were used for preference prediction model. The demographics of the respondents are shown in the **Figure 9c**. The gender breakdown is 174 (58%) male, 122 (41%) female, and one person did not respond. In terms of age, the 25-34 age group had the most participants with 106 (35%), followed by the 35-44 age group with 81 (27%). There were 56 (18%) of 45-54 age group, 31 (10%) of 55-64 age group, 14 (4%) of 65-74 age group, and 9 (4%) 25 and under. In terms of house region, 140 (47%) live in the suburbs, 95 (31%) live in large cities, 54 (18%) live in small towns, and 8 (2%) live in agricultural areas.

Data preprocessing task is as follows. The ranking data was preprocessed into the choice data. The six wheels that made up one problem were compared one-on-one and set as wheels in which "high-ranked" wheels were "selected" (customer choice). This is to overcome the problem of data shortage for small survey data. As a result, 15 comparisons were generated per problem (i.e., C(6,2)), and 240 choice data can be obtained per person. Of the 16 questions, 14 were assigned as train data, 2 as validation data, and 2 as test data, respectively, so that there were no overlapping wheel designs. The train and validation data were augmented 10 times using rotation. Finally, the dataset was used at the ratio of train, validation, and test as 1,800, 300, and 30. For this case, augmentation is to prevent overfitting and build more robust models.

### 3.2.2 Individual Preference Prediction

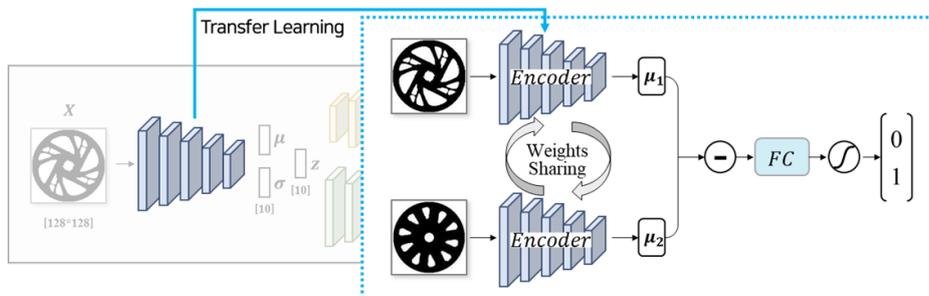

**Figure 10 Individual choice probability model architecture**

Siamese neural networks are a specific type of neural network architecture that feature two or more subnetworks that are identical in both structure and weights (Chopra et al., 2005). In other words, the subnetworks share the same architecture and weights, allowing them to learn and process data in a similar way. This design enables the Siamese network to perform tasks such as similarity matching or connection between two comparable inputs. This is made possible by the identical subnetworks that share both architecture and weights, and this network requires fewer model parameters for estimation. The **Figure 10** illustrates the architecture of the Siamese network that was employed in this study. The process of training a DL model for individual choice probability prediction involves creating a separate model for each person, based on their survey results. The trained encoder of the multimode VAE is used for TL to enhance the accuracy of the individual choice probability prediction model.

$$\Pr(U_A > U_B) = \frac{e^{U_A}}{e^{U_A} + e^{U_B}} = \frac{1}{1 + e^{-(U_A - U_B)}}$$
$$= \frac{1}{1 + e^{-(\beta^T \mu_A - \beta^T \mu_B)}} = \frac{1}{1 + e^{-\beta^T(\mu_A - \mu_B)}} \quad (2)$$

$$\text{Utility} = \beta \times \text{attribute} + \varepsilon. \quad (3)$$

$$\text{Sigmoid}(x) = \frac{1}{1 + e^{-x}} \quad (4)$$

It is noteworthy in the structure of the model that the model architecture was constructed based on an equation that calculates the choice probability using utility in conjoint analysis (Eq 2). First, in the Eq 3, the pre-trained encoder derives the attributes of the design. The error term $\varepsilon$ replaced the unobservable factor that could affect the level of utility for each individual design. However, in DL model, we conduct TL for the encoder individually to overcome the error term about the unknown factors and newly defining individually-recognized attributes. In addition, the individual part-worth $\beta$ is replaced by the weight of the FC layer to fit the part-worth individually. In the process of obtaining attributes (latent vectors) through a VAE, $z$ changes every time depending on the $\varepsilon$ drawn from the standard distribution because z is sampled using the $\mu$, $\sigma$, and standard distribution rather than reducing the dimension immediately like AE. To eliminate this uncertainty, this study used $\mu$ as the attributes of the designs rather than $z$.

When calculating the choice probability, multiplying the attributes by part-worth results in utility. By subtracting utility of two designs and applying sigmoid activation (Eq 4), the choice probability can be obtained. When sigmoid is applied to the FC layer of the model, the DL model itself becomes the same as the structure for obtaining the choice probability. For individual model training, 297 models corresponding to total respondents were trained. Each model subtracts the two $\mu$ values obtained by inputting two image inputs into one encoder sharing the weights. And then, it passes through the FC layer and sigmoid activation in turn. We used binary cross entry (BCE) loss for training.

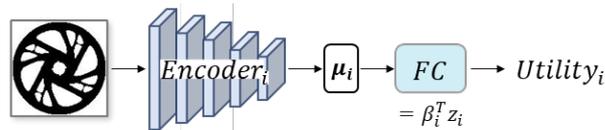

**Figure 11 Individual utility model architecture**

After completing the training of the choice probability prediction model, a utility ($U$) may be obtained immediately by utility prediction model. Utility prediction model can be defined by removing the sigmoid activation function and attaching the encoder and the FC layer (see **Figure 11**). The model weights are from choice probability prediction model, and do not need to be trained. This allows the model to predict an individual's utility for the design, which, unlike the choice probability, outputs as a continuous value.

### 3.2.3 Ensemble Preference Prediction

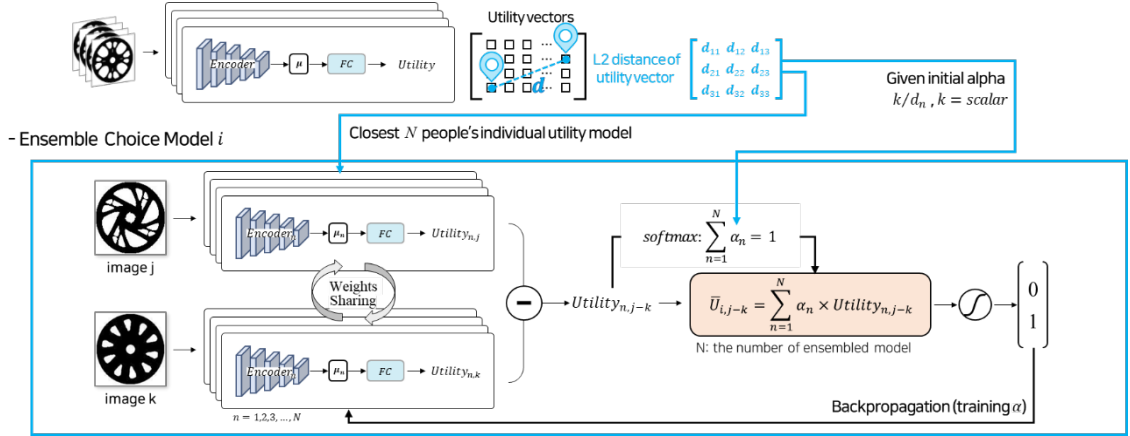

**Figure 12 Ensemble choice probability model architecture**

The individual model relied only on a small amount of individual preference tastes. Because only the non-linearity of the small amount of data is considered, the probability of overfitting is high, and there is no information about the tendency of the human subjectivity, so models can be biased. To overcome this, we introduce a method of ensemble model: ensemble choice probability model and ensemble utility model, same purpose as individual model. The ensemble choice probability model can learn one's individual preferences in consideration of public preference trends that are similar to his own together. We expect this model to prevent overfitting and result in better performance of preference prediction than individual model by understanding other people's data. The architecture of the model is shown in **Figure 12**.

1. Define Utility Vector: we predict all 297 individuals' utilities of train, validation datasets (the sum of train and validation datasets is 1,143). This makes a 297 by 1,143 shaped matrix. The utilities of 1,143 design samples were gathered from each individual and minmax scaling was applied so that the utilities for all design have values between 0 and 1. These utilities can be considered as the utility vectors of the individuals as the gathered vectors represent their preference tendencies.
2. Select Utility Models for Ensemble: In order to predict the choice probability of a person $i$, we ensemble individual utility models. To select utility model for ensemble, we calculate L2 distance of utility vectors, getting 297 by 297 distance matrix which is symmetric. Because all utility values are scaled, the values of the distance matrix stay within a region of similar values without outliers. We select 100 people that are closest to a person $i$.
3. Set Weights Alpha: The reciprocals of L2 distance in utility vectors are given as initial values of alpha $\alpha_p$ where p is 1 to 100, not equal to $i$ ($p = 1,2,...,100, p \neq i$) and the initial value of $\alpha_i$ itself is set between 0.1-10. In order not to deviate alpha from the reasonable range, the softmax activation function is applied to alpha so that the sum of alpha always becomes 1. The weights were also set to trainable parameters.
4. Ensemble Individual Models: Ensemble choice probability model is also trained individually. Same as individual model, Siamese network is used. We predict utilities $Utility_j$ and $Utility_k$ of two design samples $j$ and $k$, and subtract them ($Utility_{j-k}$). By repeating this task for 100 individual models respectively, we get $Utility_{p,j-k}$, where p is 1 to 100. Weight alpha $\alpha_p$ is multiplied for each $Utility_{p,j-k}$ to get $\overline{U}_{i,j-k}$, ensembled $Utility_{i,j-k}$ for individual $i$. Thereafter, the choice probability is derived by applying sigmoid activation function to $\overline{U}_{i,j-k}$.
5. Update Alpha: While training, we freeze the model weights of the encoder and the FC layer, only updating the individual's weights alpha through backpropagation. Alpha will be fitted in the direction of minimizing individual's model loss. It may be the direction of where the weight of people with similar preferences to those of individual increases, and in the direction where the weight of people with opposite preferences decreases. At this time, we can learn the weights and optimize which person's model will further influence the preference prediction. The ensemble model's optimized alpha value helps to minimize prediction errors and ensure the highest possible accuracy in the predicted utilities.

The individual choice probability model is trained with only 72 of the 96 wheels answered by the respondent, excluding validation, test data set. Even if the augmentation task is carried out, this is still the case that data is insufficient. The number of data is far from enough to predict the choice for the new design. We overcome this limitation of individual model through ensemble method. All surveys consisted of questionnaires that combined various designs, and were all different from one another. Using the model of a person with similar preferences to individual, interpolation is also possible for design samples that the respondent has not seen.

### 3.3 Design Recommendation System

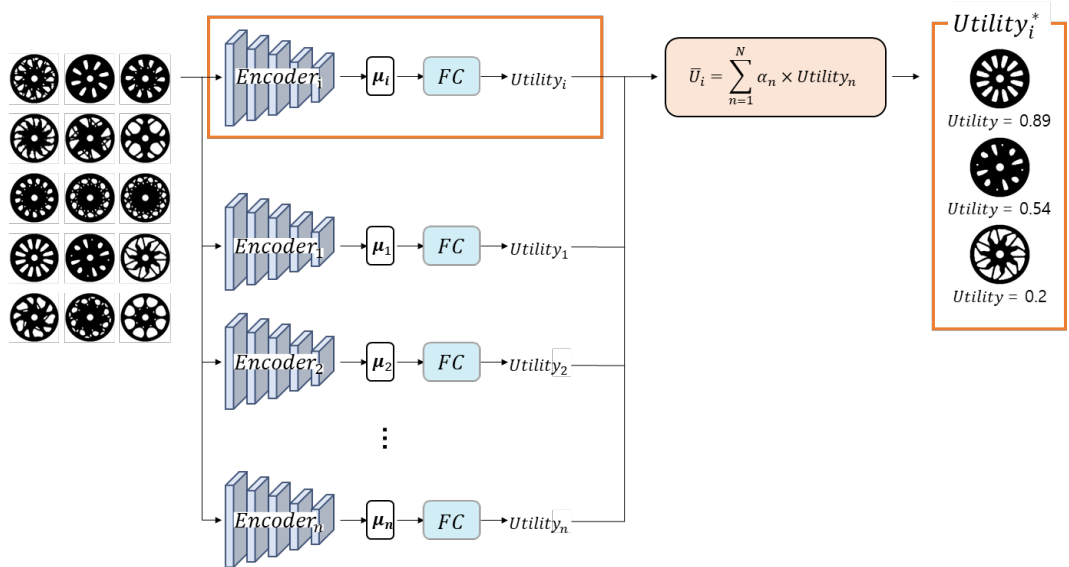

**Figure 13 Design recommendation using ensemble utility model**

    This study has developed a recommendation system using wheel design as an example, which employs ensemble utility models to predict utility and recommend designs with the highest utility to individuals. The same process as depicted in the **Figure 13** is followed for predicting utilities. By removing sigmoid activation function from ensemble choice probability model, ensemble utility model is defined. Ensemble utility models that incorporate optimized alpha values are used to improve the accuracy of the predictions. The proposed recommendation system is useful in predicting heterogeneous utility, which allows for the determination of whether a new design is meaningful to individuals. By recommending designs with high utility, this system can assist individuals in making decisions and can facilitate the design process.

# 4. Results and Discussion

## 4.1 Multi-modal VAE

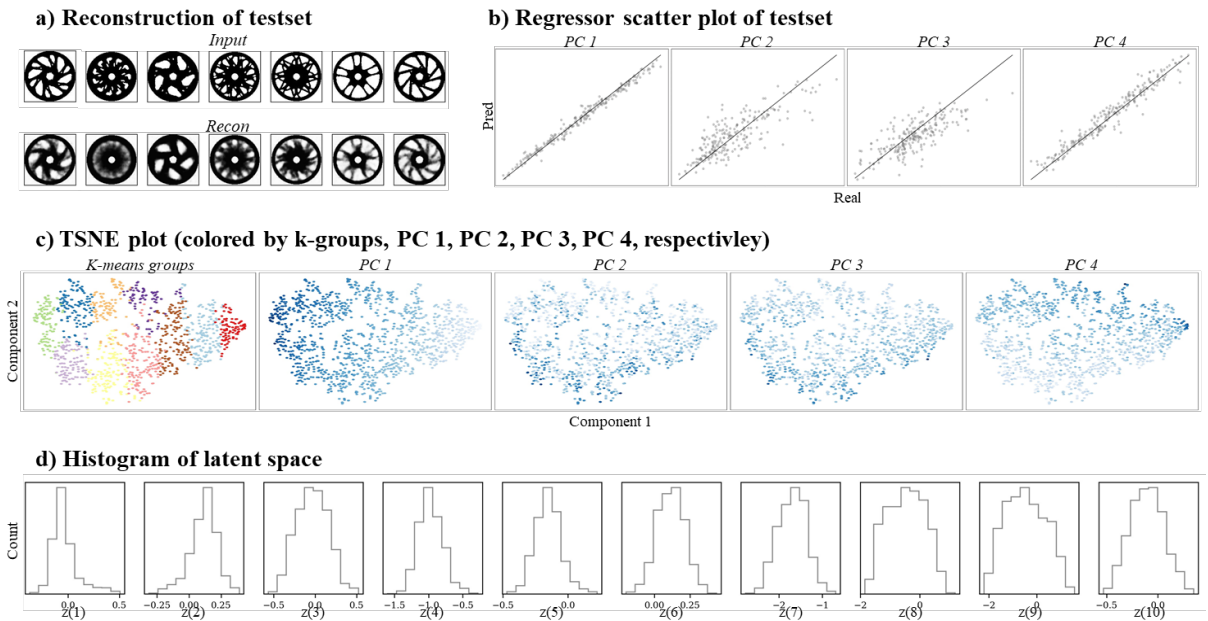

**Figure 14 Multi-modal VAE training results**

Mutlimodal VAE was trained with a multi-loss problem to utilize encoder. The training result was the same as the **Figure 14**. **Figure 14a** is the shape reconstructed by the decoder, and **Figure 14b** is the scatter plot of labels (PC1~4) and predicted values. The reconstruction might not be better than that of vanilla VAE as it is a multi-loss problem. But still, it can be judged that both shape approximation and label prediction were performed well enough. To make sure that the encoder has a good grasp of the features, we clustered the latent vectors of 1,156 images into 10 groups through k-means clustering, and visualized them using TSNE (see **Figure 14c**). When colored according to the groups, it was confirmed that the components of each group were closely gathered and well clustered apart from other groups. In addition, we colored the TSNE plot with the PCs to see if the encoder trained the PCs well (see **Figure 14c**). This also confirms that the latent vectors are linearly distributed for each PC, proving that the encoder approximates the PCs well. We could also check latent variable for 10 dimensions by the histogram of latent space (see **Figure 14d**).

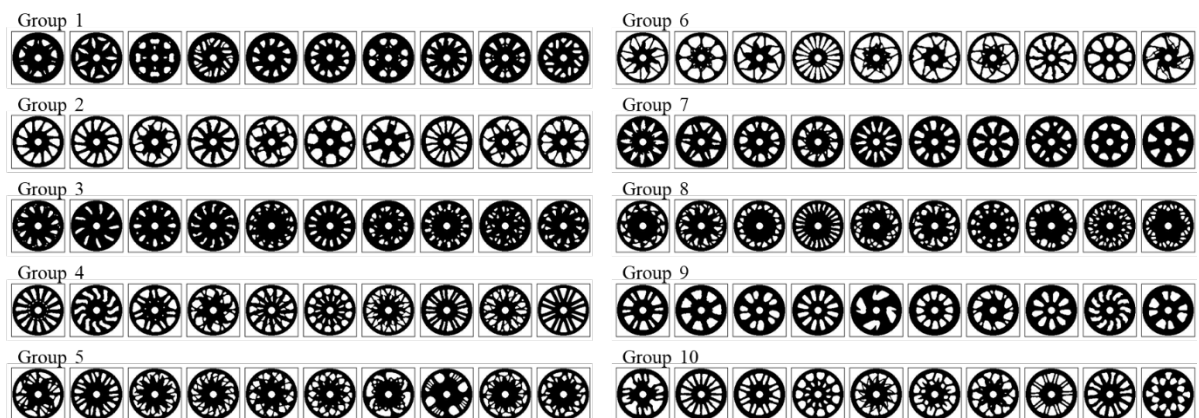

**Figure 15 Wheel design visualization according to the clustered groups**

In order to confirm that vectors with similar features are located closely in the latent space, 10 designs of each group that are closest from the centroid, were printed in the order of the distance from the centriod (see **Figure** 15). This showed that the encoder learned the shape well and that the latent vectors of similar shapes were located close to each other. The model took approximately 56.78 minutes to train with one GPU (Geforce RTX 3090).

## 4.2 Choice Probability Prediction

**Table 3 Accuracy for choice probability prediction**

| VAE Model | Preference Prediction Model | Accuracy | |
| --- | --- | --- | --- |
| | | Mean | Median |
| vanilla VAE | HB (vae vector) | 55.71% | 56.66% |
| | Population choice probability model | 50.00% | - |
| | Individual choice probability model | 58.63% | 60.00% |
| multi-modal VAE | HB (vae vector) | 59.18% | 60.00% |
| | Population choice probability model | 50.00% | - |
| | Individual choice probability model | 61.50% | 63.33% |
| | **Ensemble model w/ trainable alpha** | 64.75% | 66.67% |

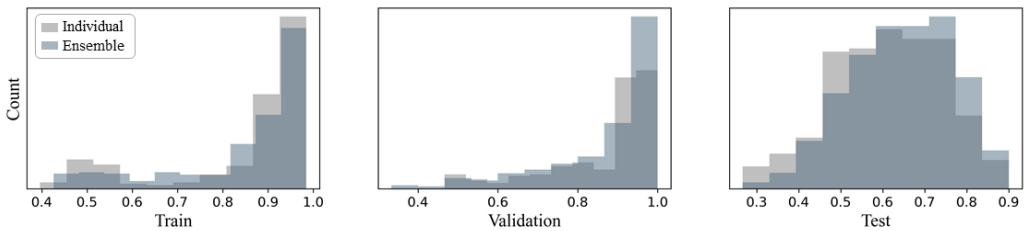

**Figure 16 Accuracy histogram of ensemble choice model with trainable alpha**

The accuracy of the surrogate models for choice probability is given in **Table 3**. It was compared with the accuracy of the test dataset which is equally given to all surveys. For the compassions, HB using vae vectors as design attributes, population choice probability model, and individual choice probability model were compared to determine the validity of our proposed model.

First, the most widely used HB technique for individual preference modeling in marketing field was computed using the ChoiceModelR library in R language (Sermas and Colias, 2012). HB is computed using the latent vector, which is dimensionally reduced to 10 dimensions through the encoder of VAE / multi-modal VAE. The part-worth $\beta$ value obtained by HB is multiplied by latent vector to calculate the personal utility value. Calculating the choice probability based on the utility to find the accuracy of the test data, we get an accuracy of 55.71% using the vanilla VAE vector and 59.18% using the multi-modal VAE vector. The calculation time of HB was 67.92 and 68.27minutes, respectively.

Second, the population choice probability model has the same architecture as the individual choice probability model, but instead of the individual model, it assumes a homogeneous model where all people have a same distribution, and then trains by combining all data. Inferring the individual test datasets together gives an accuracy of 50.00% for both the vanilla VAE and multi-modal VAE. In the binary classification problem, the convergence of the accuracy to 0.5 proves that predicting homogeneous level preferences is meaningless.

Third, the prediction of choice probability through the individual model introduced in section 3.2.2 can obtain an accuracy of 58.63% and 61.50% for the vanilla VAE and multi-modal VAE, respectively. The training time was between 14 ~ 15 seconds for each individual, and it takes about 12 minutes when computing in parallel with 6 people on 3 GPUs (Geforce RTX 3090). Finally, our proposed model, ensemble model with trainable alpha, achieves 64.75% accuracy, the highest for choice probability prediction. The ensemble training itself takes about 7~8 seconds for each individual and about 6.3 minutes for the entire ensemble with the same equipment setting as the individual.

**Figure 16** shows three histogram plots of 297 people for train, validation, and test data respectively. We see that the weighted ensemble method has shifted the accuracy of the models to the higher side overall.

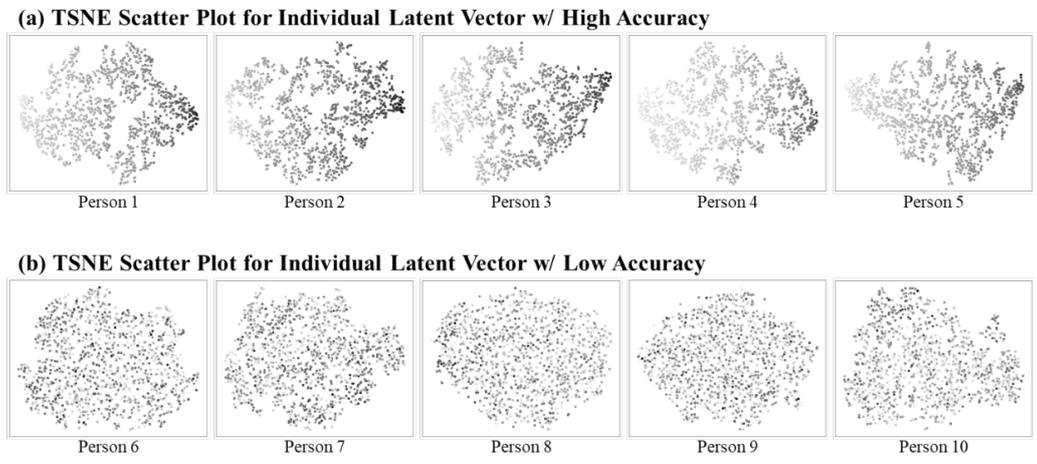

Figure 17 TSNE scatter plot for individual latent vectors

The 10 individual latent vectors were obtained by passing the design samples through individual encoders: five for the individuals with an accuracy of 0.8 or higher and five for individuals with an accuracy of 0.5 or lower on the ensemble model. Representing the latent vectors in a 2D space with TSNE and coloring it with the utilities predicted by the ensemble model is shown in the **Figure 17**. We can see that for models with high accuracy in predicting choice probabilities, the latent space is fitted to understand and reflect personal utility (see **Figure 17a**, person 1~5). However, for the less accurate models, we see that the latent vectors do not capture the trend of the utilities and are cluttered (see **Figure 17b**, person 6~10).

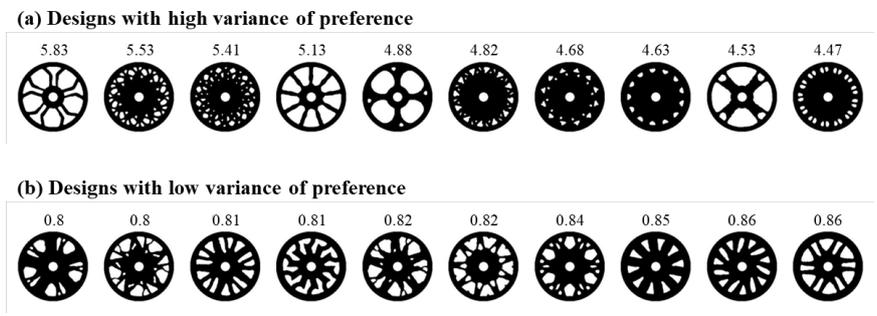

Figure 18 Wheel designs sorted by variance of preference

To see which designs had the most influence on people's preferences, we calculated the variance of the 297 people's design preferences for each wheel and identified the 10 wheels with the highest and 10 with the lowest variance (**Figure 18**). A design with a large variance can have a large impact on people's design preferences because it indicates a wide range of preferences. In **Figure 18**, designs with features such as thin or very thick spokes, a very clean or a very complex, were found to have high variance. It can be determined that people have a stronger "like" or "dislike" emotion for a design with more prominent features than for a design with less descriptive features. In future studies, it may be possible to obtain more meaningful results by designing surveys that are more likely to influence people's preferences.

## 4.3 Design Recommendation

### 4.3.1 Design Recommendations for Individuals

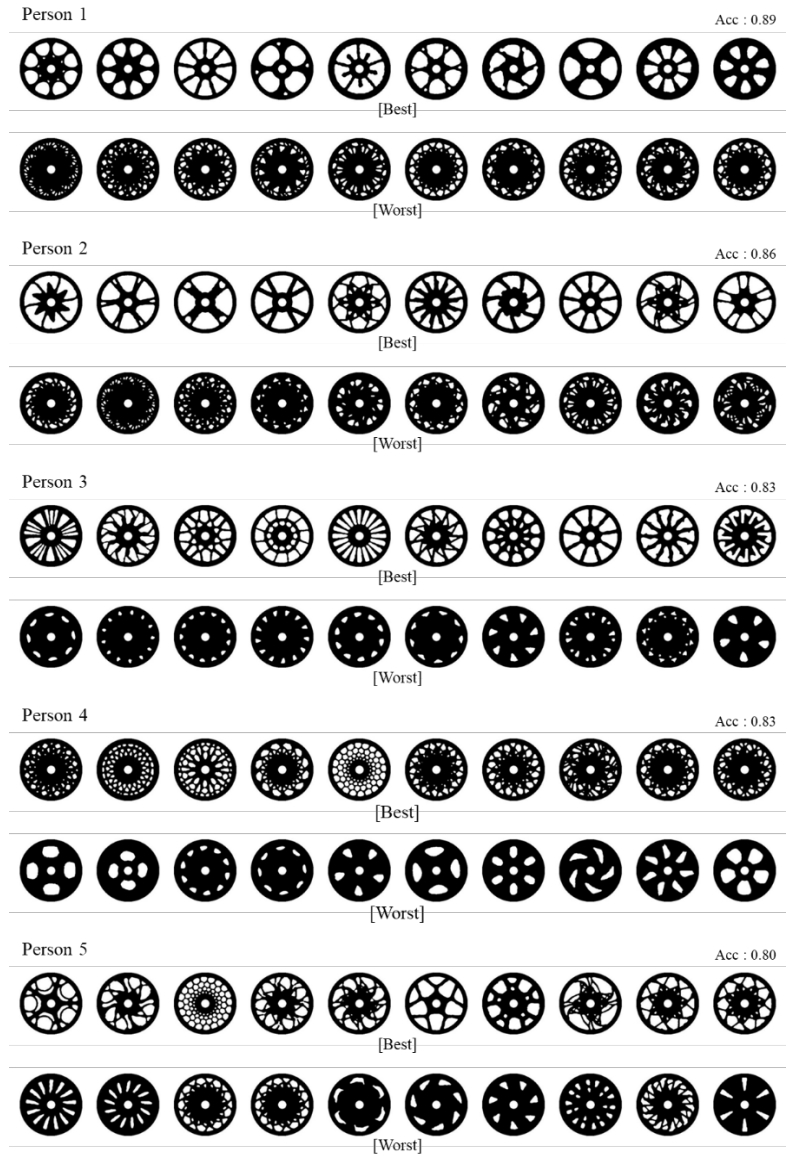

**Figure 19 Best/worst design for 5 individuals w/ highest accuracy**

To verify that the model has learned the tendencies of heterogeneous preferences well, we checked the best/worst wheel designs for the five people with the highest accuracy for the test dataset (see **Figure 20**). Person 1 showed an overall preference for curved, non-angled spoke geometry. He preferred wheels with fewer spokes to wheels with more spokes, and also tended to prefer wheels with spokes that were not densely clustered in the center of the wheel. Person 2 preferred a straight, stretched designs with thin spokes. Neither person 1 or 2 preferred complex designs with a lot of tiny spheres.

Person 3 tended to prefer wheels with a bicycle wheel-like design, with thin but many spokes, and didn't preferred dull, simple wheels. Person 4 preferred dense, ornate, spoked wheels, with a preference that was completely different from the population's preference. He disliked a thick, blunt, curved design with fewer spokes. Lastly, person 5 preferred a splendid and aesthetic designs, with spokes extending into twigs.

### 4.3.2 Design Recommendations for Groups

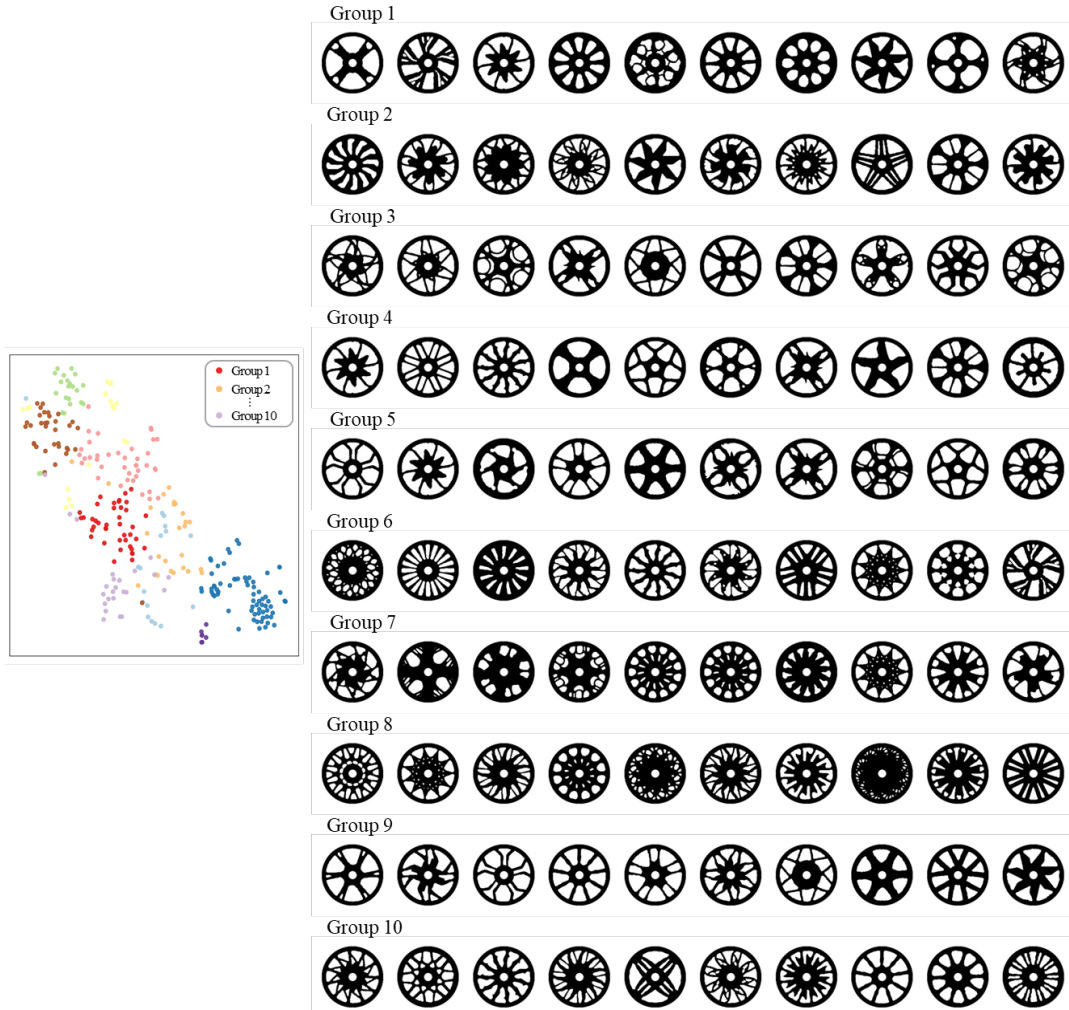

**Figure 20 Clustering customers' utility vectors**

Assuming that there are groups of people with similar preferences, we dimensionally reduced the utility vector to a 10-dimensional vector with AE and expressed it in two dimensions with TSNE to understand the preferences of each group. The 10-dimensional reduced vector was clustered by k-means clustering and the TSNE scatterplot shown in **Figure 21**(left-side) was colored with the groups. The number of groups was set to 10, which is where the elbow point occurred after calculating the k-means elbow method. Group 1-10 were separately well distributed in 2D space, which means that there are some distinguishing characteristic of utility vectors. Comparing the best wheels for each group, we can see the preference tendency of each group as shown in **Figure 21**(right-side). The design tendencies of groups are not as discrete as those of the multi-modal VAE (**Figure 15**), but each group has its own tendencies. For instance, Group 4,5, and 9 preferred designs with simple and thin spokes, Group 7 and 8 seemed to prefer most ornate designs. Groups 1 and 3 preferred designs with relatively fewer than six spokes, while Group 6 preferred wheels with relatively more spokes.

### 4.3.3 Design Recommendations for Demographics

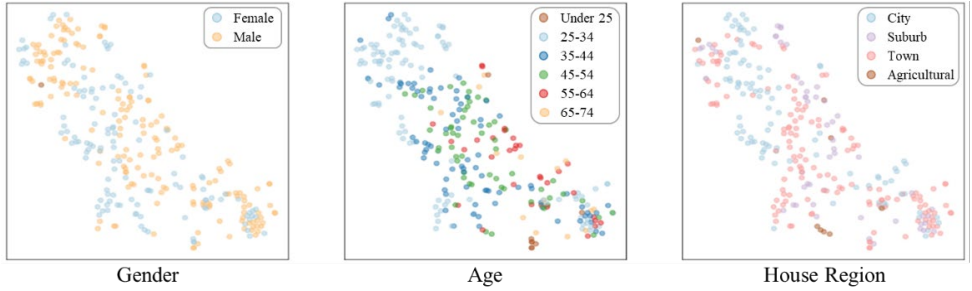

Figure 21 TSNE plots colored by demographics

Linking customer's demographics and preferences is essential for making design recommendations for target customer groups as well as individual customers. For unknown customers with no preference information, a design recommendation system can be generalized if it can recommend preferences to customers based on their basic demographics. Therefore, it should be possible to approximate the utility function of an unknown customer with demographics about the customer (e.g. consumer label in Pan et al. (2017)). In this research, we did not use customer information for predicting preferences, we have not conducted surveys or built frameworks that take demographics into account, so a new experimental design is needed to account for them. But to see the scalability of our framework, we tried to check the correlation between demographic information and utility in the survey. We visualized the previous AE dimensionality-reduced vectors using demographic information (see **Figure 22**). Each color corresponds to a category mentioned in **Figure 9c**. However, we could not find some special correlation between them.

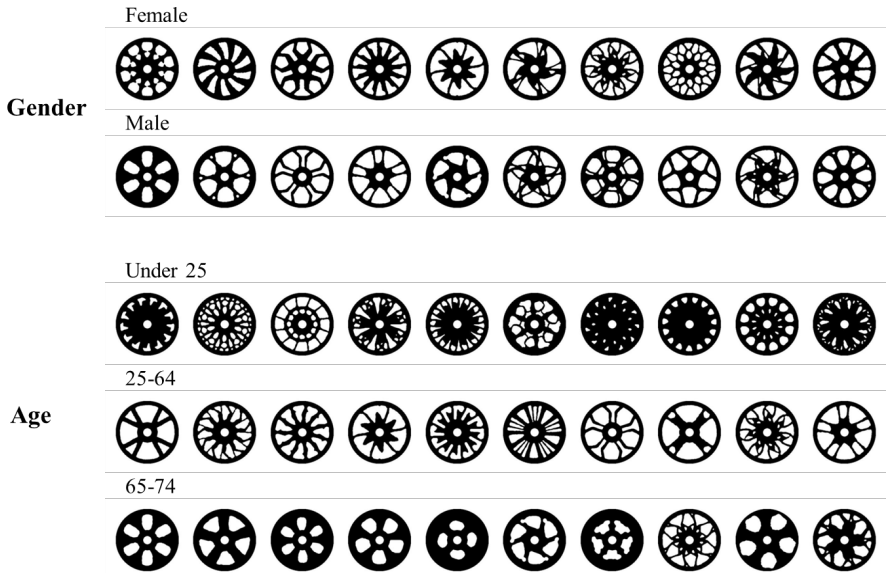

Figure 22 Preferred designs based on demographic categories

For each of the three demographic categories, gender, age, and house region, the designs were ordered by the averaged utility (see **Figure 23**). As we saw in the TSNE scatterplot, the utilities did not show a strong tendency toward the demographic information, but there was a trend when we list the actual design samples. comparing male and female, Female preferred branched and ornate wheels more than male. Men preferred simpler wheels with larger openings. When comparing by age group, the 25 and under group preferred a patterned design with small openings and lots of small spheres, while the four groups between the ages of 25 and 64 had similar visual preferences for straight, large-opening wheels. Those aged 65 and older preferred simpler designs with thicker spokes on the rim part to give them a duller feel. As mentioned earlier, further analysis of design preferences with customer information would require an experimental design using customer information. This is essential future work for the scalability of the preference prediction problem.

# 5. Conclusion

This study proposes a DL-based design preference model that predicts heterogeneous design preferences suggesting a design representation method to obtain features that simultaneously reflect the designs and the designer's domain knowledge and an ensemble method based on individual preference similarity that can improve predictive power with less data.

Firstly, we train multi-modal VAE to extract features. This model is capable of feature extraction that simultaneously considers the features of the design shape itself and the features reflecting the designer's domain knowledge. Second, we proceed a preference prediction modeling, using Siamese network and weighted ensemble method. We train individual preference model, which can predict individual utilities. By using this individual utility model, we calculate utility vectors of individuals and ensemble them under the consideration of similarity of preferences. Lastly, we can suggest design recommendation system, based on the ensemble preference model.

We have built a framework to improve three key factors of preference prediction: design representation, preference modeling, and heterogeneity. The contribution of this study to the above task is as follows.
1. We construct a latent space that understands and learns the parameters defined by human. This allows to automatically reduce the dimensionality for new designs to latent vectors, while considering the defined features.
2. We solve the problem of individual-level model which can be biased by limited data and models by using ensemble method to optimize the model.
3. We suggest recommendation system that enables companies to determine design concepts for target markets, analyze products for trade-offs between product performance and preference, and provide customized design recommendations to consumers.

We demonstrated that the high dimensionality of the designs, the non-linearity of preferences, and the limitations of individual-level models can be improved through design representation and ensemble techniques. However, this study has several limitations.

First, as mentioned earlier, the methodology needs to be extended to match customer information and recommend preferences. To this end, research can be conducted on surveying various customer groups, selecting meaningful data from the survey data that is representative of the population, and building a system that can predict utility using customer information.

Second, other product attributes need to be considered. In order to design a product, not only aesthetics, but also technical factors such as engineering performance and manufacturability must be considered in a variety of ways. When the trade-offs between them are properly considered, a marketable product can be developed. In the future, a product development process that considers the customer's preferences for designs, engineering performance, and various technical factors at the same time is needed.

Third, it should be able to reflect real-time customer feedback (Kang et al., 2019). In our framework, models are trained with surveyed data and re-training is needed to update with new data. However, customer feedback is updated in real time and must be considered to build an effective design recommendation system. Therefore, design recommendation system using real-time training such as ML, active learning, and reinforcement learning should be suggested to update the model with customer feedback in real time.

# References

[1] Bloch, P. H. (1995). Seeking the ideal form: Product design and consumer response. Journal of marketing, 59(3), 16-29.
[2] Bloch, P. H., Brunel, F. F., & Arnold, T. J. (2003). Individual differences in the centrality of visual product aesthetics: Concept and measurement. Journal of consumer research, 29(4), 551-565.
[3] Burnap, A., Hauser, J. R., & Timoshenko, A. (2023). Product aesthetic design: A machine learning augmentation. Marketing Science.
[4] Chopra, S., Hadsell, R., & LeCun, Y. (2005, June). Learning a similarity metric discriminatively, with application to face verification. In 2005 IEEE computer society conference on computer vision and pattern recognition (CVPR'05) (Vol. 1, pp. 539-546). IEEE.
[5] Dotson, J. P., Beltramo, M. A., Feit, E. M., & Smith, R. C. (2019). Modeling the Effect of Images on Product Choices. Available at SSRN 2282570.
[6] Dunteman, G. H. (1989). Principal components analysis (No. 69). Sage.
[7] Evgeniou, T., Boussios, C., & Zacharia, G. (2005). Generalized robust conjoint estimation. Marketing Science, 24(3), 415-429.
[8] Fenech, C., & Perkins, B. (2019). The Deloitte Consumer Review. Made-to-Order: The Rise of Mass Personalisation. Deloitte Development LLC. https://www2.deloitte.com/content/dam/Deloitte/ch/Documents/consumer-business/ch-en-consumer-business-made-to-order-consumer-review.pdf
[9] Gabel, S., & Timoshenko, A. (2022). Product choice with large assortments: A scalable deep-learning model. Management Science, 68(3), 1808-1827.
[10] Gensch, D. H., & Recker, W. W. (1979). The multinomial, multiattribute logit choice model. Journal of Marketing Research, 16(1), 124-132.
[11] Golanty, J. (1997). Conjoint analysis and choice modeling considerations. Marketing Research, 9(1), 4.
[12] Green, P. E., & Srinivasan, V. (1990). Conjoint analysis in marketing: new developments with implications for research and practice. Journal of marketing, 54(4), 3-19.
[13] Green, P. E., Krieger, A. M., & Wind, Y. (2004). Thirty years of conjoint analysis: Reflections and prospects (pp. 117-139). Springer US.
[14] Johnson, R. M. (2000). Understanding HB: An intuitive approach. Sawtooth Software Inc., Sequim, WA.
[15] Kang, N., Ren, Y., Feinberg, F., & Papalambros, P. (2019). Form+ function: Optimizing aesthetic product design via adaptive, geometrized preference elicitation. arXiv preprint arXiv:1912.05047.
[16] Kelly, J. C., Maheut, P., Petiot, J. F., & Papalambros, P. Y. (2011). Incorporating user shape preference in engineering design optimisation. Journal of Engineering Design, 22(9), 627-650.
[17] Kotler, P., & Alexander Rath, G. (1984). Design: A powerful but neglected strategic tool. Journal of business strategy, 5(2), 16-21.
[18] Landwehr, J. R., Labroo, A. A., & Herrmann, A. (2011). Gut liking for the ordinary: Incorporating design fluency improves automobile sales forecasts. Marketing Science, 30(3), 416-429.
[19] Lee, S. (2020). Deep Learning-based Design Preference Prediction and Recommendation System: A Case Study of Car Wheel Design, M.S. Thesis, Department of Mechanical Systems Engineering, Sookmyung Women's University.
[20] Lenk, P. J., DeSarbo, W. S., Green, P. E., & Young, M. R. (1996). Hierarchical Bayes conjoint analysis: recovery of partworth heterogeneity from reduced experimental designs. Marketing Science, 15(2), 173-191.
[21] Liu, J., Zhi, Q., Ji, H., Li, B., & Lei, S. (2021). Wheel hub customization with an interactive artificial immune algorithm. Journal of Intelligent Manufacturing, 32, 1305-1322.
[22] Liu, Y., Li, K. J., Chen, H., & Balachander, S. (2017). The effects of products' aesthetic design on demand and marketing-mix effectiveness: The role of segment prototypicality and brand consistency. Journal of Marketing, 81(1), 83-102.
[23] Louviere, J. J., & Woodworth, G. (1983). Design and analysis of simulated consumer choice or allocation experiments: an approach based on aggregate data. Journal of marketing research, 20(4), 350-367.
[24] Lugo, J. E., Batill, S. M., & Carlson, L. (2012, August). Modeling product form preference using Gestalt principles, semantic space, and Kansei. In International Design Engineering Technical Conferences and Computers and Information in Engineering Conference (Vol. 45066, pp. 529-539). American Society of Mechanical Engineers.
[25] Mturk. (2020). Retrieved from https://www.mturk.com/
[26] Oh, S., Jung, Y., Kim, S., Lee, I., & Kang, N. (2019). Deep generative design: Integration of topology optimization and generative models. Journal of Mechanical Design, 141(11).

[27] Orsborn, S., Cagan, J., & Boatwright, P. (2009). Quantifying aesthetic form preference in a utility function.
[28] Pan, Y., Burnap, A., Hartley, J., Gonzalez, R., & Papalambros, P. Y. (2017, August). Deep design: Product aesthetics for heterogeneous markets. In Proceedings of the 23rd ACM SIGKDD International Conference on Knowledge Discovery and Data Mining (pp. 1961-1970).
[29] Reid, T. N., MacDonald, E. F., & Du, P. (2013). Impact of product design representation on customer judgment. Journal of Mechanical Design, 135(9), 091008.
[30] Rossi, P. E., & Allenby, G. M. (2003). Bayesian statistics and marketing. Marketing Science, 22(3), 304-328.
[31] Ryu, N., Seo, M., & Min, S. (2021). Multi-objective topology optimization incorporating an adaptive weighed-sum method and a configuration-based clustering scheme. Computer Methods in Applied Mechanics and Engineering, 385, 114015.
[32] Sermas, R., & Colias, M. J. V. (2012). Package 'ChoiceModelR'.
[33] Sisodia, A., Burnap, A., & Kumar, V. (2022). Automatically Discovering Unknown Product Attributes Impacting Consumer Preferences.
[34] Sylcott, B., Michalek, J. J., & Cagan, J. (2013, August). Towards understanding the role of interaction effects in visual conjoint analysis. In International Design Engineering Technical Conferences and Computers and Information in Engineering Conference (Vol. 55881, p. V03AT03A012). American Society of Mechanical Engineers.
[35] Treiber, B., & Needel, S. P. (2000). New WAYS TO EXPLORE CONSUMER POINT-OF-PURCHASE DYNAMICS. The Impact of Networking: Marketing Relationships in the, 263.
[36] Tseng, I., Cagan, J., & Kotovsky, K. (2012). Concurrent optimization of computationally learned stylistic form and functional goals.
[37] Veryzer Jr, R. W., & Hutchinson, J. W. (1998). The influence of unity and prototypicality on aesthetic responses to new product designs. Journal of consumer research, 24(4), 374-394.
[38] Walker, J., & Ben-Akiva, M. (2002). Generalized random utility model. Mathematical social sciences, 43(3), 303-343.
[39] Yoo, S., Lee, S., Kim, S., Hwang, K. H., Park, J. H., & Kang, N. (2021). Integrating deep learning into CAD/CAE system: generative design and evaluation of 3D conceptual wheel. Structural and multidisciplinary optimization, 64(4), 2725-2747.